\documentclass[preprint,preprintnumbers,amsmath,amssymb,superscriptaddress]{revtex4-1}
\usepackage{graphicx}
\usepackage{multirow}
\usepackage{dcolumn}
\usepackage{bm}
\usepackage[usenames,dvipsnames]{xcolor}

\usepackage{amsmath}

\usepackage{mathtools}

\usepackage{ulem}
\definecolor{darkgreen}{rgb}{0,0.6,0}

\newcommand{\diff} [1]{\mathrm{d}{#1}}                  % differential in the integral
\newcommand{\Vphi}{\vv{\phi}}                           % phase-field vector (order parameter)
\newcommand{\phia}{\phi_{\alpha}}                       % phase field phi_alpha
\newcommand{\phib}{\phi_{\beta}}                      % phase field phi_alpha
\newcommand{\phid}{\phi_{\delta}}                        % phase field phi_beta

\newcommand{\vv}[1]{\boldsymbol{#1}}
\newcommand{\calF}{\mathcal{F}}

\newcommand{\gabd}{\gamma_{\alpha \beta\delta}}

\newcommand{\rhs}{\text{lhs}}

\begin{document}
\preprint{ \color{NavyBlue} version 2.0 - \today}

\title{\Large \color{NavyBlue}  Compound droplets on fibers}

\author{Floriane Weyer}
\affiliation{GRASP, Physics Department, University of Li\`ege, 4000 Li\`ege, Belgium }
\author{Marouen Ben Said}
\affiliation{Institute of Applied Materials - Computational Materials Science, Karlsruhe Institute of Technology, 76131 Karlsruhe, Germany}
\author{Johannes H\"{o}tzer}
\affiliation{Institute of Materials and Processes, University of Applied Sciences, 76133 Karlsruhe, Germany}
\author{Marco Berghoff}
\affiliation{Institute of Applied Materials - Computational Materials Science, Karlsruhe Institute of Technology, 76131 Karlsruhe, Germany}
\author{Laurent Dreesen}
\affiliation{GRASP, Physics Department, University of Li\`ege, 4000 Li\`ege, Belgium }
\author{Britta Nestler}
\affiliation{Institute of Applied Materials - Computational Materials Science, Karlsruhe Institute of Technology, 76131 Karlsruhe, Germany}
\affiliation{Institute of Materials and Processes, University of Applied Sciences, 76133 Karlsruhe, Germany}
\author{Nicolas Vandewalle}
\affiliation{GRASP, Physics Department, University of Li\`ege, 4000 Li\`ege, Belgium }
\thanks{Correspondance : GRASP, Physics Department B5a, University of Li\`ege, B-4000 Li\`ege, Belgium. http://www.grasp-lab.org }

%%%%%%%%%%%%%%%%%%%%%%%%%%%%%%%%%%%%%%%%%%% ABSTRACT
\begin{abstract}
{\bf 

Droplets on fibers have been extensively studied in the recent years. Although the equilibrium shapes of simple droplets on fibers are well established, the situation becomes more complex for compound fluidic systems. Through experimental and numerical investigations, we show herein that compound droplets can be formed on fibers and that they adopt specific geometries. We focus on the various contact lines formed at the meeting of the different phases and we study their equilibrium state. It appears that, depending on the surface tensions, the triple contact lines can remain separate or merge together and form quadruple lines. The nature of the contact lines influences the behavior of the compound droplets on fibers. Indeed, both experimental and numerical results show that, during the detachment process, depending on whether the contact lines are triple or quadruple, the characteristic length is the inner droplet radius or the fiber radius.}
\end{abstract}
\maketitle
%%%%%%%%%%%%%%%%%%%%%%%%%%%%%%%%%%%%%%%%%%% INTRODUCTION
\section{Introduction}
In the recent years, many studies have focused on droplets hanging on horizontal fibers. Such systems are very different from sessile droplets. Indeed, the curvature of the substrate induces some typical phenomena such as the inhibition of film formation \cite{bro, mac1, mac2}. This means that even if the spreading coefficient is greater than zero, droplets can be formed on fibers. Moreover, the geometry of such systems has been investigated leading to the definition of two equilibrium configurations of a droplet on a fiber : the barrel and the clamshell shapes \cite{mac3,carr1,wu,carr2}. The barrel shape is easy to characterize mathematically thanks to its symmetry \cite{carr1}. However, a numerical approach is necessary for the clamshell shape. The public domain software Surface Evolve (SE), developed by Brakke \cite{brakke}, has been established for numerical simulations of various wetting problems, especially for studying the equilibrium morphologies of single droplets on cylindrical fibers \cite {mac3,ruiter,chou}. The transition between both equilibrium geometries has also been analyzed \cite{carr2,mac3}. A droplet adopts a barrel configuration when its volume is large or when the radius of the fiber is small. When the volume decreases or the radius increases, the droplet changes its shape and adopts the clamshell shape. This transition also depends on the history of the system. Moreover, electrowetting has been used to change the properties of the droplets hanging from fibers and therefore to provoke the transition \cite{ruiter, eral, dufour}. The influence of gravity has also been investigated \cite{ruiter, chou}.

Recent works \cite{gileta, giletb} demonstrated that droplets on fibers may constitute the starting point of an open digital microfluidics, emphasizing the interest for those systems. Moreover, optofluidic devices can be designed by using optical fiber networks \cite{lismont}. However, the major difficulties of such devices are the contamination and the evaporation of the droplets. This can be slowed down by the encapsulation of the core water droplets into an oil shell leading to the creation of compound droplets.
From a numerical point of view, the situation becomes more complex for systems containing several components. At our knowledge, such wetting problems can not be numerically solved by the SE software. Therefore, in a recent work \cite{said}, we introduced a multiphase-field model which enables the numerical investigation of multi-droplet systems, in particular of compound droplets. In this framework, each phase refers to a droplet with specific physical properties, such as density and surface tension. The temporal evolution of each phase occurs in order to minimize the free energy of the system. 

The main objective of the present paper is the study of such compound droplets on fibers. Horizontal fibers will be used in order to analyze such systems in their equilibrium state. This is a first step towards the elaboration of digital microfluidics devices on fibers. First, we focus on the encapsulation process and we determine the conditions under which a water droplet is encapsulated by an oil droplet. Second, we show that, when a compound droplet is placed in a substrate, different contact lines are formed at the junctions between the various phases. These triple contact lines can be stable or move and fuse together. In this case, a quadruple line is formed. Third, we briefly introduce the phase-field model that we use as numerical support to confirm our experimental results. In the last two sections, we focus on compound droplets on fibers. In particular, we investigate the detachment process which highlights the existence of a quadruple line.  

%%%%%%%%%%%%%%%%%%%%%%%%%%%% DROPLET ENCAPSULATION
\vskip 0.4cm
\section{Experimental}

In this study, we investigate the case of compound droplets made of water and oil.  We used not only pure  water ($w$) but also soapy water ($s$) for the core droplets in order to study the influence of the surface tension. Pure water is colored in blue to get stronger contrast whereas soapy water is colored in red. Soapy water is a mixture of pure water and dish-washing liquid based on anionic surfactant AEOS-2EO \cite{dreft}. Both liquids have a density $\rho _ w = \rho _s =1000 \ \rm{kg/m^3}$. The oil is a silicone oil with a viscosity of 20 cSt and a density $\rho _ o = 949 \ \rm{kg/m^3}$. For the different fluids encountered in this study, the surface tensions were measured through pendant drop method using a CAM 200 goniometer from KSV Instruments Ltd \cite{cam}. Each value is the average of five different experiments consisting in taking twenty pictures of a pendant drop and deducing the surface tension from the shape of the droplet. The surface tensions are found to be $\gamma_{A/w} = 71.1 \pm 0.2$ N/
m, $\gamma_{A/s}$  = 25.7 $\pm$ 0.1 N/m, $\gamma_{A/o}$  = 18.9 $\pm$ 1.4 N/m, $\gamma_{o/w}$ = 40.1 $\pm$ 1.9 N/m and $\gamma_{o/s}$ = 4.6 $\pm$ 0.1 N/m. 

\section{Results and discussion}
\subsection{Droplet encapsulation}

\begin{figure}[h]
\begin{center}
\vskip 0.2 cm
\includegraphics[width=0.6\textwidth]{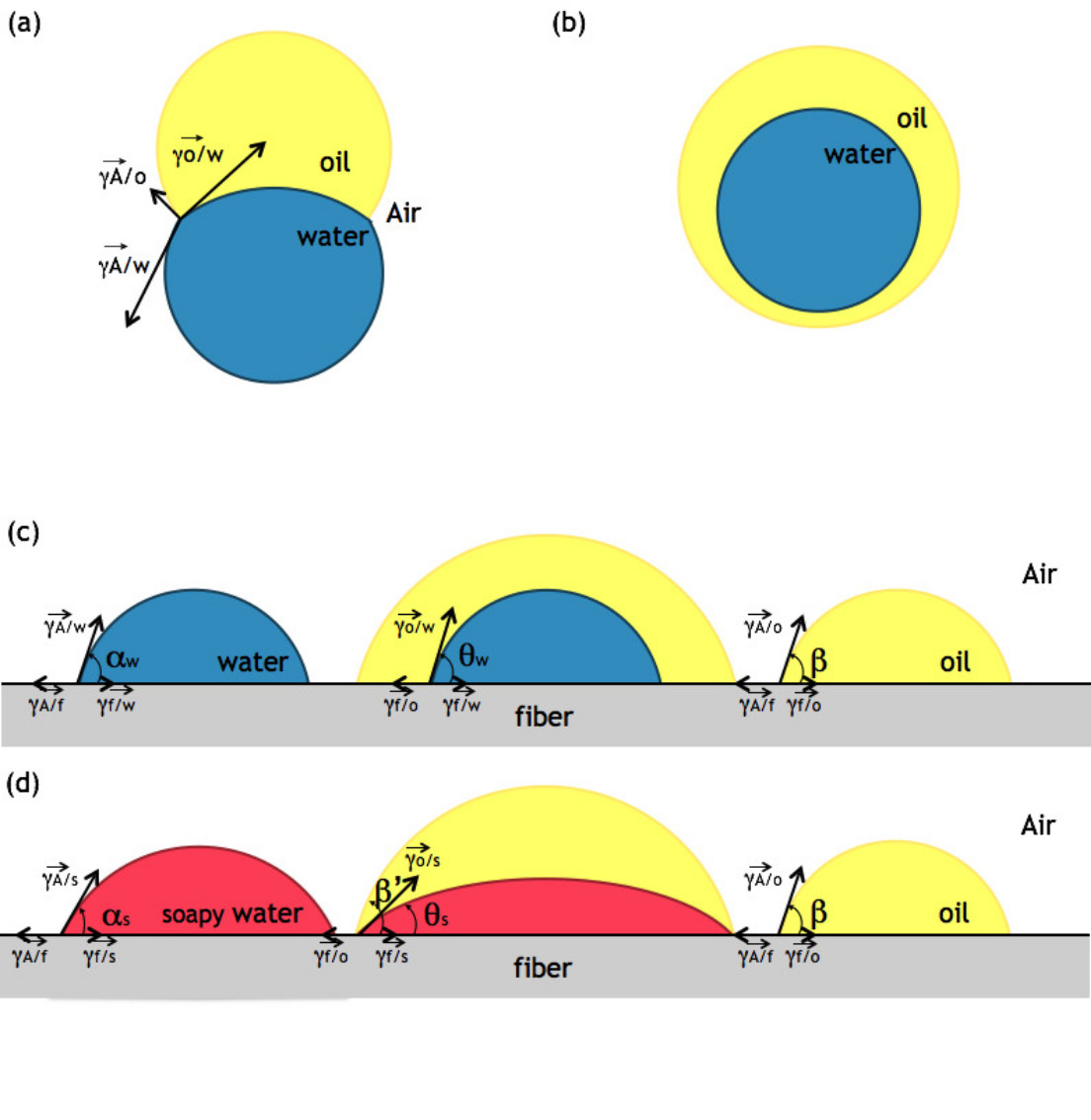}
\vskip -0.2 cm
\caption{ (a) Sketch of a compound droplet in air. Three tensions compete at the triple junction. (b) Sketch of the compound droplet for which the condition of Equation (1) is satisfied : water is encapsulated by oil. (c) Sessile droplets on a planar solid substrate (denoted $f$ for fiber). At the left, a pure water droplet is shown. In the middle, there is a compound droplet. At the right, an oil droplet is in equilibrium. Three different contact lines are possible and are characterized by different contact angles : $\alpha_w$ for pure water in air, $\beta$ for oil in air, and $\theta_w$ for water in oil. Tensions are illustrated at the triple junctions.(d) The same sketch for sessile droplets containing soapy water. In that case, due to the quadruple line, the contact angle for oil containing soapy water is $\beta '$ instead of $\beta$.  }
\label{fig_sketch}
\end{center}
\end{figure}

The first step is to determine the condition under which the water is encapsulated by oil. Therefore, we first consider the fluid phases only, without any substrates. Droplet encapsulation is provided by the balance of the surface tensions \cite{guz}. Figure \ref{fig_sketch}(a) presents the contact between two non-miscible liquids in air. A contact line is formed where three surface tensions compete. The balance of these tensions defines the contact angle of the compound system. Encapsulation of water by an oil shell is possible when the air-water tension ($\gamma_{A/w}$) is larger than the sum of two other tensions ($\gamma_{A/o}$ for the air-oil interface and $\gamma_{o/w}$ for the oil-water interface). The condition 
\begin{equation}
\gamma_{A/w} > \gamma_{A/o} + \gamma_{o/w}
\label{cond}
\end{equation}should be satisfied, otherwise the oil droplet is unable to cover the water core. The condition is the same for soapy water. In view of the surface tensions, the above condition (\ref{cond}) is fulfilled for both water and soapy water and therefore encapsulation is expected as shown in Figure \ref{fig_sketch}(b).

%%%%%%%%%%%%%%%%%%%%%%%%%%%%%%%%%%%%%%%%%%% SOLID PLANAR SUBSTRATE
\vskip 0.4cm
\subsection{Encapsulation on a substrate}
If we consider a compound droplet on a planar solid substrate (denoted $f$ because a planar substrate can be seen as a fiber with an infinite radius), the situation may become tricky \cite{neeson} and simulations are often used in order to predict the geometry of the system \cite{said}. Assuming complete encapsulation of the water droplet by oil, as discussed in the previous section, two contact lines are present on a sessile compound droplet as pictured in Figure \ref{fig_sketch}(c). The inner contact line is the boundary between oil, water and the substrate and is characterized by a contact angle $\theta_w$. The outer contact line is at the meeting of air, oil and the substrate, and defines a contact angle $\beta$. We can compare this situation with the cases of a water and an oil sessile droplet. For a water droplet on a planar surface, there is only one contact line (at the meeting of air, water and the surface) characterized by a contact angle $\alpha_w$ which differs from $\theta_w$. For a sessile oil 
droplet, the unique contact line is characterized by the same contact angle $\beta$ as defined previously as shown in Figure \ref{fig_sketch}(c). The equilibrium condition for each type of contact line on a planar substrate is given by the Young's equation and leads to the following system
\begin{eqnarray}
\gamma_{f/o} & = & \gamma_{f/w} + \gamma_{o/w} \cos \theta_{w} \\
\gamma_{A/f} & = & \gamma_{f/w} + \gamma_{A/w} \cos \alpha_{w} \\
\gamma_{A/f} & = & \gamma_{f/o} + \gamma_{A/o} \cos \beta 
\end{eqnarray}
which can be combined in \begin{equation}
\cos \theta_{w} = {\gamma_{A/w} \cos \alpha_{w} - \gamma_{A/o} \cos \beta \over \gamma_{o/w}}.
\label{cos}
\end{equation}
A similar expression can be written for the case of soapy water described in Figure \ref{fig_sketch}(d).

In order to predict the value of $\cos \theta_w$ ($\cos \theta_s$), the contact angles $\alpha_w$ ($\alpha_s$) and $\beta$ were measured. As the contact angles only depend on the surface tensions according to the Young's equation, the contact angles should be the same on a plane or on a fiber \cite{quere,liang, liang2}. Thus, contact angles were measured when pure water, soapy water and oil droplets are hanging on nylon fibers. For each contact angle, ten pictures were taken of droplets hanging on fibers. We found that $\alpha_w = 58.9 ^{\circ} \pm 10.3 ^{\circ}$ for a sessile water droplet. However, it appears that for both soapy water and oil, the contact angles are smaller :  $\alpha_s = 20.7 ^{\circ} \pm 7.8 ^{\circ}$ and $\beta = 24.0 ^{\circ} \pm 7.7 ^{\circ}$, respectively. By considering the values of surface tensions and contact angles, it is possible to evaluate the right member of Equation (\ref{cos}) and then to estimate the contact angle formed by the water drop when it is covered by oil. For pure water encapsulated by oil, the value of $\theta_w$ is equal to $61.0^{\circ}\pm 16.4 ^{\circ}$ meaning that the water droplet forms a spherical cap with a well-defined contact angle surrounded by oil. In this case, there are two distinct contact lines : the first one is the junction of oil, water and substrate and the second one is the meeting of air, oil and substrate.
Whereas, for soapy water, the evaluation of the right member of Equation (\ref{cos}) leads to a value greater than 1 (namely 1.5). This means that $\theta_s$ is non valued. In this case, soapy water spreads inside the oil drop. The inner contact line moves towards the outer contact line. Depending on the soapy water and the oil volumes, the contact lines could either never meet or merge together. In the latter case, a quadruple line could be formed. This quadruple line represents the union of four phases which are air, soapy water, oil and fiber. Even if a quadruple line is formed, the observation of this phenomenon is very difficult as both soapy water and oil spread and meet the surface with small contact angles. It has to be noticed that the quadruple line formation may modify the contact angle of the oil shell. Moreover, the existence of this quadruple line has been theoretically predicted by Mahadevan \textit{et al.} \cite{mahadevan} even if this assumption is against the thermodynamics laws. Indeed, thermodynamics suggests that three phases meet along a line whereas four phases should join at a single point. Prior to the present work, this quadruple line has never been observed experimentally, at our knowledge. 

%%%%%%%%%%%%%%%%%%%%%%%%%%%%%%%%%%%%%%%%%%% Numerical Method
\vskip 0.4cm

\subsection{Numerical Method}

Before considering the case of compound droplets on a fiber, we introduce the numerical method used to support the experimental results.
When a droplet is put in contact with a solid surface, its shape evolves in order to adopt the energetically most favorable morphology. Mathematically, this process can be seen as a minimization problem with evolving free boundaries. Diffuse interface methods, in particular phase-field methods have been proved as thermodynamically consistent and powerful numerical tools to investigate such type of problems \cite{anderson, jacqmin, ding,nakamura}. The simulation results in the present work are based on the model presented in \cite{said}, extended by an energy term $f_{gv}$ reflecting the gravitational effect. The free energy of the system reads
\begin{align}
	\calF (\Vphi)= \int_{\Omega} \left(\epsilon a (\Vphi, \nabla \Vphi) + \frac{1}{\epsilon} w (\Vphi) + g(\Vphi) + f_{gv}(\Vphi) \right) \diff {\Omega}+ \int_{\partial_s \Omega} f_w(\Vphi)
	\, \diff S.
	\label{eq:ginzbur_landau_functional}
\end{align}
The spatial domain is denoted by $\Omega$ and its solid boundary (here the fiber) by $\partial_s \Omega$. $\epsilon$ is a small positive parameter, related to the thickness of the diffuse interface.
Here, $\Vphi(\mathbf{x},t)= \left(\phi_w(\mathbf{x},t), \phi_o(\mathbf{x},t), \phi_A(\mathbf{x},t)\right)$ is the so-called phase-field vector, where each component $\phia(\mathbf{x},t)$, $\alpha \in\{ w, o, A\}$ describes the state of the phase $\alpha$ in time and space. The subscripts $w, o$ and $A$  refer to the water, oil and air phases, respectively. When considering soapy water instead of water, the subscript $s$ is used. Additionally, we postulate the phasefield variables $\phia$ sum up to unity in each point of space, as well as $0\leq\phia\leq 1$. So, $\phia$ can be interpreted as the volume fraction of the phase $\alpha$ and variates continuously along the diffuse interface between $0$ and $1$. The first term in the free energy functional Equation \eqref{eq:ginzbur_landau_functional} is a gradient energy
density expressed as
% density expressed, in terms of a generalized gradient verctors $\qab = \phia \nabla \phib - \phib \nabla \phib$, as
\begin{equation}
	a(\Vphi, \nabla \Vphi) = \sum_{\alpha < \beta} \gamma_{\alpha \beta} |\phi_{\alpha}\nabla\phi_{\beta}  - \phi_{\beta} \nabla\phi_{\alpha}|^2.
	\label{eq:afunc}
\end{equation}
The second addend in Equation \eqref{eq:ginzbur_landau_functional} is a higher order multi-obstacle potential, of the form
 \begin{equation}
	w(\phi) = \begin{cases}
							\frac{16}{\pi^2}\sum_{\alpha < \beta} \gamma_{\alpha \beta}\phi_{\alpha} \phi_{\beta} + \sum_{\alpha < \beta<\delta}\gabd \phia 
							\phib \phid, &\mbox{if } \phi \in [0,1], \\
							\infty &\mbox{else}.
							\label{eq:func_w}
						\end{cases}
\end{equation}
The bulk energy contribution
		\begin{equation}
			g(\Vphi) = \sum_{\alpha} \chi_{\alpha} h(\phi_{\alpha})
				\label{eq:bulk_energ}
		\end{equation}
		ensures the volume preservation of each phase in the system. For more details, we refer to \cite{said} and the references therein.
		The energy contribution $f_{gv}(\Vphi)$ reflects the gravitational effect and reads
				\begin{equation}
			f_{gv}(\Vphi) = \sum_{\alpha} \rho_{\alpha} \mathbf{g} \cdot \mathbf{x} h(\phia)
				\label{eq:bulk_energ}
		\end{equation}
		where $\bf{g}$ represents the gravity force, $\rho_{\alpha}$ the density of the corresponding phase, $\bf x$ the position in the computational domain and $h(\phi) = \phi_{\alpha}^3(6 \phi_{\alpha}^2-15 \phi_{\alpha}+10)$ interpolates the phasefield values along the diffuse interface.
		The surface integral in Equation \eqref{eq:ginzbur_landau_functional} models the interaction between the liquid droplets and the fiber and is responsible for the establishment of the correct contact angles at the fiber:
		
				\begin{align}
			f_w(\Vphi) = \sum_{\alpha=1}^N\gamma_{\alpha f} h(\phia) + m \sum\limits_{\alpha<\beta<\delta }^ N \phia \phib \phi_{\delta}.
				\label{eq:fw}
		\end{align}
		Here, $\gamma_{\alpha f}$ are the surface tensions between the phase $\alpha$ and the fiber, for each phase $\alpha\in \{w, o, A\}$. The parameter $m$ as well as $\gabd$ (in Equation \eqref{eq:func_w}) are phenomenological parameters discussed in \cite{said, garcke, stinner} and will be given later for each numerical simulation.

		The evolution equations for each phase $\alpha$ in the system is determined by minimizing the free energy functional \eqref{eq:ginzbur_landau_functional} using methods of variational
		calculus. A steepest descent method is applied, conducting to the following system of time dependent evolution equations
			\begin{align}
			\underbracket{\epsilon (\nabla \cdot a_{,\nabla\phia}(\phi,\nabla\phi) -a_{,\phia}(\phi,\nabla\phi)) - \frac{1}{\epsilon} w_{,\phia}(\phi) - g_{,\phia}(\phi) + {f_{gv}}_{,\phia}(\phi)}_{=:\rhs_{1}} -
			\lambda_1 = \tau \epsilon \partial_t \phia  \quad \text{in} \; \Omega
				\label{eq:pf_and_bc0}
		\end{align}
		with the natural boundary condition
		\begin{align}
			\underbracket{ - \epsilon a_{,\nabla\phia}(\phi,\nabla\phi)\cdot\vv {n} - f_{w,\phia}(\phi)}_{=:\rhs_{2}} -\lambda_2 & = 0 \quad \text{on} \; \partial_S\Omega .
				\label{eq:pf_and_bc}
		\end{align}
		The time relaxation parameter $\tau$ is set to unity in all simulations. The notation $a_{,\nabla\phi_{\alpha}}$, $a_{,\phi_{\alpha}}$, $w_{,\phi_{\alpha}}$, $g_{,\phi_{\alpha}}$ and $f_{w,\phia}$ is used to indicate the partial derivatives $\partial/\partial_{\nabla\phi_{\alpha}}$ and $\partial/\partial \phi_{\alpha}$ of the functions $a (\Vphi, \nabla \Vphi)$, $w (\Vphi)$,
		$g(\Vphi)$ and $ f_w(\Vphi)$, respectively. The divergence of the vector field $a_{,\nabla\phi_{\alpha}}(\Vphi,\nabla\Vphi) $ is denoted by $\nabla \cdot ()$ and the time derivative $\partial \phi_{\alpha}
		(\mathbf{x}, t)/ \partial t $ is denoted by $\partial_t \phi_{\alpha}$. 
		The normal to the fiber, $\partial_s\Omega$, is denoted by $\vv{n}$. $\lambda_1$ and $\lambda_2$ are Lagrange multipliers, according to the constraint $\sum_{\alpha} \phia =1$.
		\begin{align}
			\lambda_1 = \frac{1}{N^*}\sum_{\alpha =1}^ {N^*} \rhs_1, \quad  \lambda_2 = \frac{1}{N^*}\sum_{\alpha =1}^{N^*} \rhs_2 
				\label{eq:lamda1_lambda2}
		\end{align}
		The evolution equations (3)  are discretized using finite differences on a regular grid with equidistant spacings in all spatial directions. For the discretisation in time, the forward Euler scheme was applied, which requires the control of the time step $\Delta t$ for ensuring numerical stability.

%%%%%%%%%%%%%%%%%%%%%%%%%%%%%%%%%%%%%%%%%%% FIBER
\vskip 0.4cm
\subsection{Encapsulation on a fiber}

Let us focus on compound droplets on fibers. Is it possible to create and observe compound droplets and quadruple lines on a fiber? From the information obtained in the previous sections, we can assume that, on a fiber, the water droplet should be encapsulated by oil. Moreover, for pure water, two different contact lines should be formed whereas, for soapy water, a quadruple line should be created. However, the wetting behaviors on a planar substrate are significantly different from those on a cylindrical fiber. First, droplets on fibers can adopt two different geometries : a clamshell shape or a barrel shape. Second, the curvature of the fiber influences the geometry of the system \cite{bro, mac1, mac2, cox1, cox2,liang,liang2}. The local curvature can inhibit the spreading of droplets. So, using a fiber could stimulate the quadruple line formation. However, these theoretical predictions have to be verified experimentally and numerically.

\begin{figure}[h]
\begin{center}
\vskip 0.2 cm
\includegraphics[width=1\textwidth]{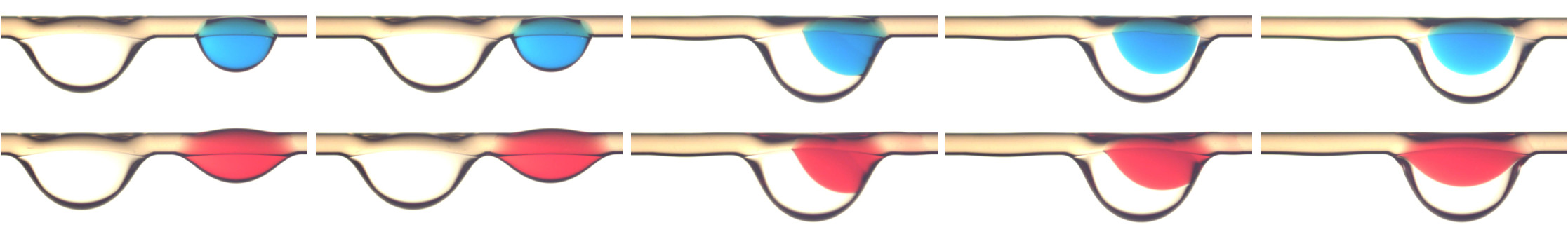}
\vskip -0.2 cm
\caption{Pictures showing the formation of compound droplets on a fiber. (Top) An oil droplet ($2.5 \ \mu l$) is placed right next to a pure water droplet ($1.0 \ \mu l$) on a $450  \ \rm{\mu m}$ diameter fiber. Due to the surface tensions, the oil progressively encapsulates the water component leading to the creation of a compound droplet. (Bottom) An oil drop ($2.5 \ \mu l$) is left next to a soapy water droplet ($1.0 \ \mu l$). The water phase is also encapsulated by oil and therefore a compound droplet is formed. The last pictures show the difference between a pure or a soapy water component : the water drop forms a well-defined core inside the oil shell, with two different contact lines on each side of the drop, whereas the soapy water component is spread along the fiber leading to the fusion of the contact lines and the possible formation of a quadruple line.}
\label{enca}
\end{center}
\end{figure}

\begin{figure}[h]
\begin{center}
\vskip 0.2 cm
\includegraphics[width=1\textwidth]{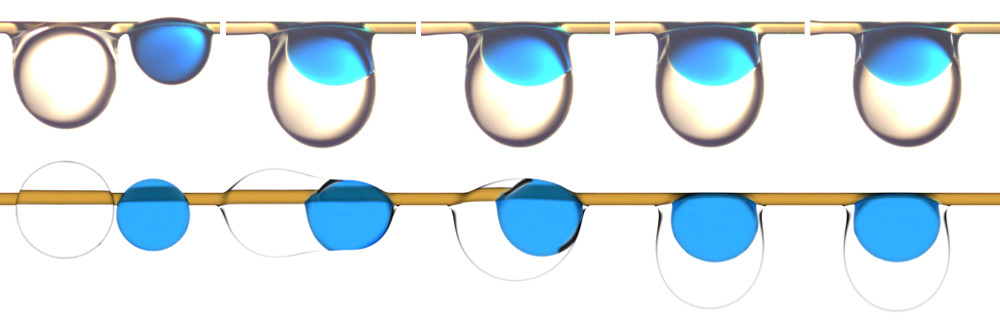}
\vskip -0.2 cm
\caption{Comparison between (top) experimental and (bottom) numerical results. The encapsulation process leads to the formation of a compound droplet ($1.0 \ \mu l$ of pure water and $2.5 \ \mu l$ of oil) on a $200  \ \rm{\mu m}$ diameter fiber. The oil progressively encapsulates the water droplet leading to the creation of a compound droplet.}
\label{encapsnum}
\end{center}
\end{figure}

Compound droplets are created by, first, placing the water drop on the fiber, thanks to a pipette, and, second, by adding oil. Figure \ref{enca} illustrates this process : (top) an oil droplet ($2.5 \ \mu l$) is placed right next to a droplet of pure water ($1.0 \ \mu l$). Due to the surface tensions, the oil droplet wraps around the water droplet. The pure water droplet adopts a  clamshell shape whereas oil forms an asymmetric barrel shape with a thin layer around the fiber. The presence of two types of contact lines is obvious. The contact line between air, oil and the fiber is separate from the oil-water-fiber line. The same situation is shown for soapy water instead of pure water. In this case, both droplets seem to adopt a barrel geometry. Moreover, the picture tends to prove the existence of a quadruple line on each side of the drop in the soapy water case. Indeed, it is extremely hard to distinguish two different contact lines.  At our knowledge, this is the first time that this type of structure seems to be created experimentally. The encapsulation process is studied numerically and compared with the experimental observations as presented in Figure \ref{encapsnum}. Starting with two spherical droplets (a pure water droplet and an oil droplet), the simulation also leads to the formation of a compound drop on a fiber. The results are visualized with the open source 3D computer graphics software Blender. A refractive index $n=1.1$ is used for the silicone oil in order to show that the triple lines are separated. The next section will evidence the formation of a quadruple line.

%%%%%%%%%%%%%%%%%%%%%%%%%%%%%%%%%%%%%%%%EQUILIBRIUM POSITION

%%%%%%%%%%%%%%%%%%%%%%%%%%%%%%%%%%%%%%%%% DETACHMENT
\vskip 0.4cm
\subsection{Droplet detachment}

\begin{figure}[h]
\begin{center}
\vskip 0.2cm
\includegraphics[width=0.4\textwidth]{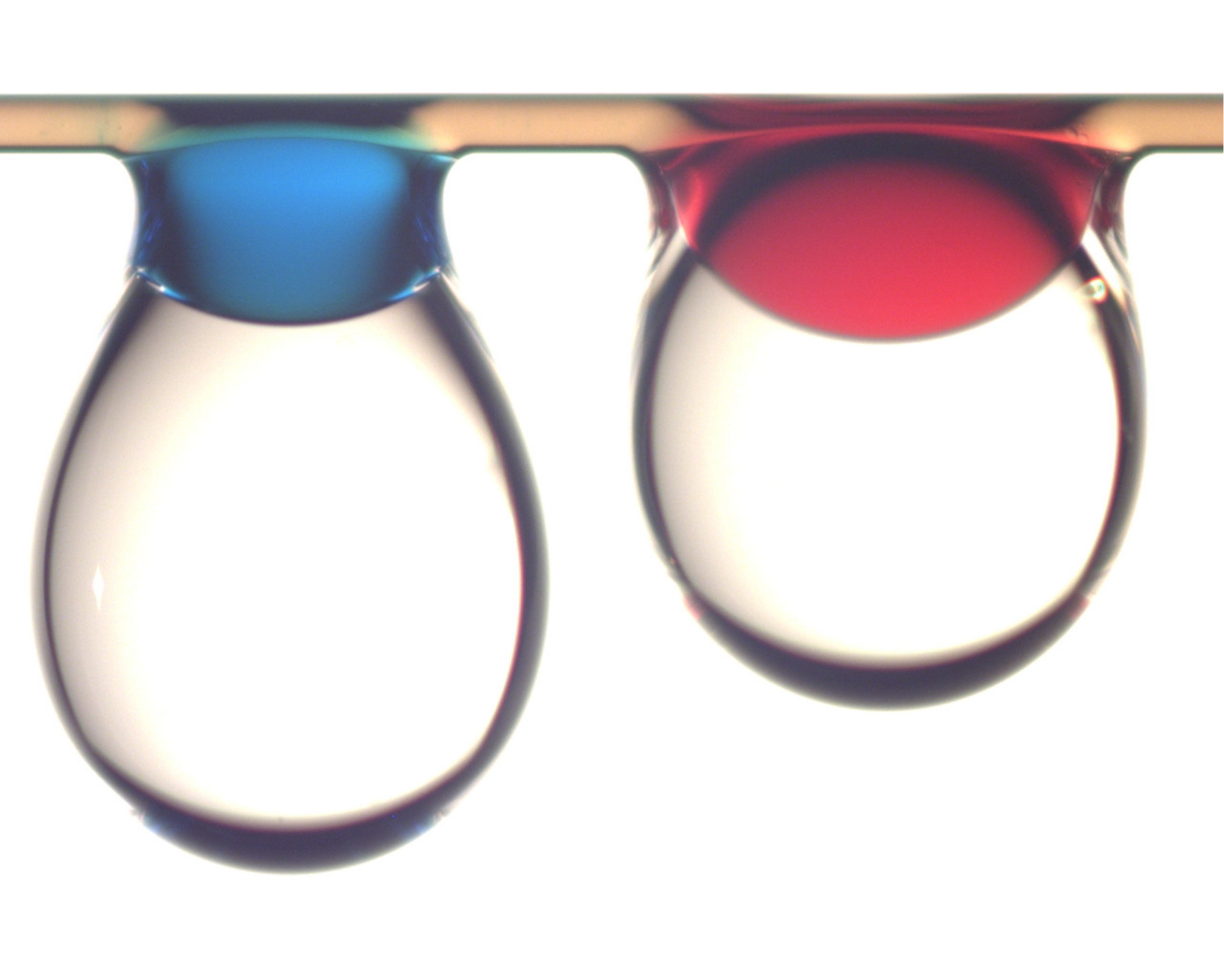}
\vskip -0.2cm
\caption{Pictures of two droplets close to detachment : (left) compound droplet with $0.5 \ \rm{ \mu l}$ of pure water and $3.9 \ \rm{\mu l}$ of oil, (right) compound droplet with $0.5 \ \rm{\mu l}$ of soapy water and $3.5 \ \rm{\mu l}$ of oil on a fiber (diameter $d=200 \ \rm{\mu m}$).  }
\label{fig_detachment}
\end{center}
\end{figure}

\begin{figure}[h]
\begin{center}
\vskip 0.2cm
\includegraphics[width=0.4\textwidth]{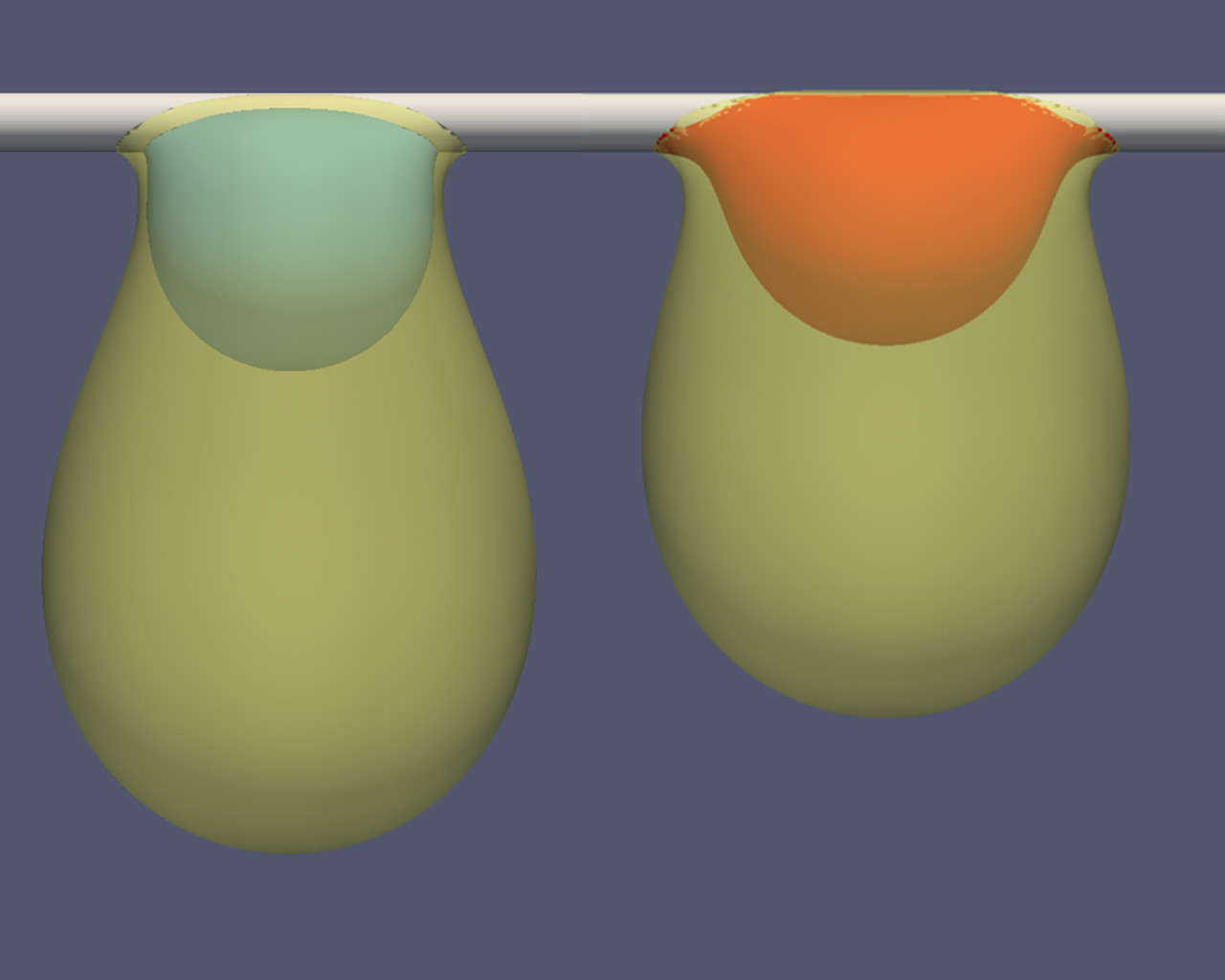}
\vskip -0.2cm
\caption{Simulation results of two droplets close to detachment : (left) compound droplet with $0.5 \ \rm{ \mu l}$ of pure water and $4.2 \ \rm{\mu l}$ of oil, (right) compound droplet with $0.5 \ \rm{\mu l}$ of soapy water and $3.5 \ \rm{\mu l}$ of oil on a fiber (diameter $d=200 \ \rm{\mu m}$).}
\label{fig_detachment_sim}
\end{center}
\end{figure}

A convincing proof of the existence of two different contact lines can be found by regarding large droplets. Figure \ref{fig_detachment} presents a picture of compound droplets close to detachment, when gravity forces start to dominate capillary forces. Both compound droplets contain $ 0.5 \ \rm{\mu l}$ of water : on the left, pure water is considered while, on the right, soapy water is shown. The volume of oil is increased gradually until the drop is ready to fall. From the experiments, we found out that the critical volume of oil is $3.9 \ \rm{\mu l}$, for pure water and $3.5 \ \rm{\mu l}$ for soapy water Figure \ref{fig_detachment}. Numerical simulations predict a critical volume of $4.2 \ \rm{\mu l}$ for pure water and $3.5 \ \rm{\mu l}$ for soapy water. Even if the oil volumes are slightly different, both experimental and numerical data show that pure water can hold larger volumes of oil than soapy water. Moreover, the shapes of the drops are different depending on the nature of the inner droplet as it can be seen from Figure \ref{fig_detachment}. The simulations also lead to these different shapes as shown in Figure \ref{fig_detachment_sim}. By comparing Figures \ref{fig_detachment} and \ref{fig_detachment_sim}, it is obvious that the experimental and numerical shapes of the compound droplets are almost identical. It has to be noticed that the shape of the inner droplet in Figure \ref{fig_detachment} is deformed because the oil shell acts as a lens and deflects the light whereas it is not the case in the simulations, which are visualized with the open source Software ParaView-4.3.1. Indeed, the light refraction is not taken into account in this software in order to visualize the shape of the inner droplet. From these figures, we can say that oil is literally hanging from the water droplet. The presence of two distinct triple lines allows the oil to move independently from the water. Whereas, in the case of soapy water, since there are only two quadruple lines, the system moves as a whole under the fiber. The shape of the system is similar to the one of a droplet only made of oil. The few differences are due to the surface tension and the contact angle that are influenced by the presence of a quadruple contact line. 


\begin{figure}[h]
\begin{center}
\vskip 0.2 cm
\includegraphics[width=0.75\textwidth]{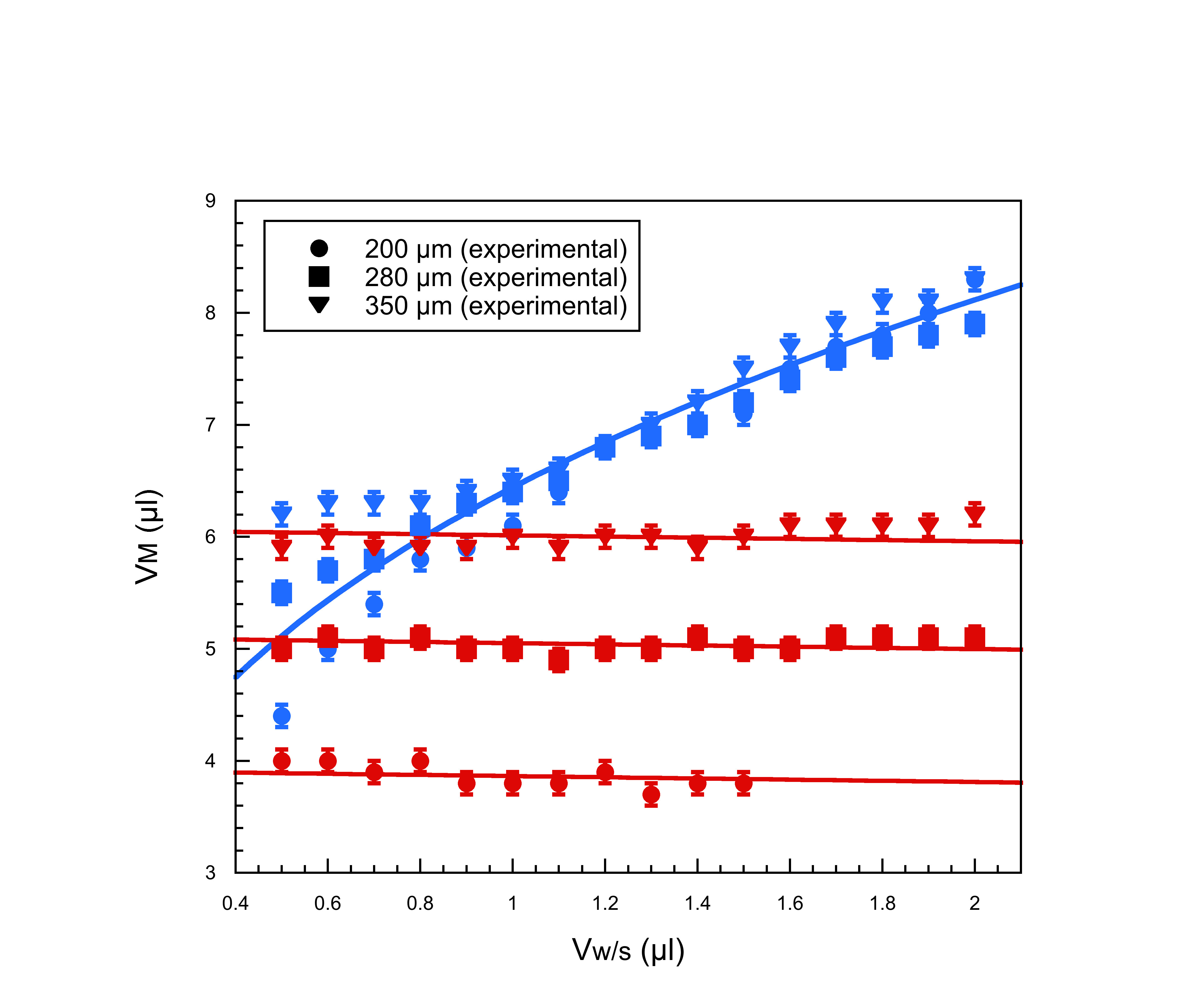}
\vskip -0.2 cm
\caption{The maximum and global volume of the compound droplet ($V_w +V_o$) that can be reached just before detachment as a function of the volume of the inner droplet. The nature of the inner droplets affects the detachment phenomenon. The pure water droplets (in blue) act like a capillary for the oil droplets and allow larger hanging droplets. The soapy water droplets (in red) do not affect the maximum volume of the whole system : the only parameter that matter is roughly the total volume. The blue curve represents the maximum volume of a droplet hanging under a capillary and given by Equation (\ref{cap}) whereas the red curve corresponds to the Equation (\ref{cap2}).}
\label{deta}
\end{center}
\end{figure}

In order to quantify this phenomenon, we reproduced this experiment for different volumes of inner droplets and for different fiber diameters ($200 \ \rm{\mu m}$, $280 \ \rm{\mu m}$ and $350 \ \rm{\mu m}$). We placed water and soapy water droplets on the fiber and we added oil progressively until the system detaches. Figure \ref{deta} presents the maximum volume, $V_M$, of the whole system (water and oil) as a function of the volume of the inner droplet, $V_{w/s}$, for three diameters. It appears that the nature of the inner droplet influences the detachment of the whole system. Indeed, for pure water, by increasing the volume of the inner droplet, the global volume is also increased. Whereas, for soapy water, the increase of the inner droplet does not affect the maximal volume which seems to be a constant. 

\begin{figure}[h]
\begin{center}
\vskip 0.2 cm
\includegraphics[width=0.75\textwidth]{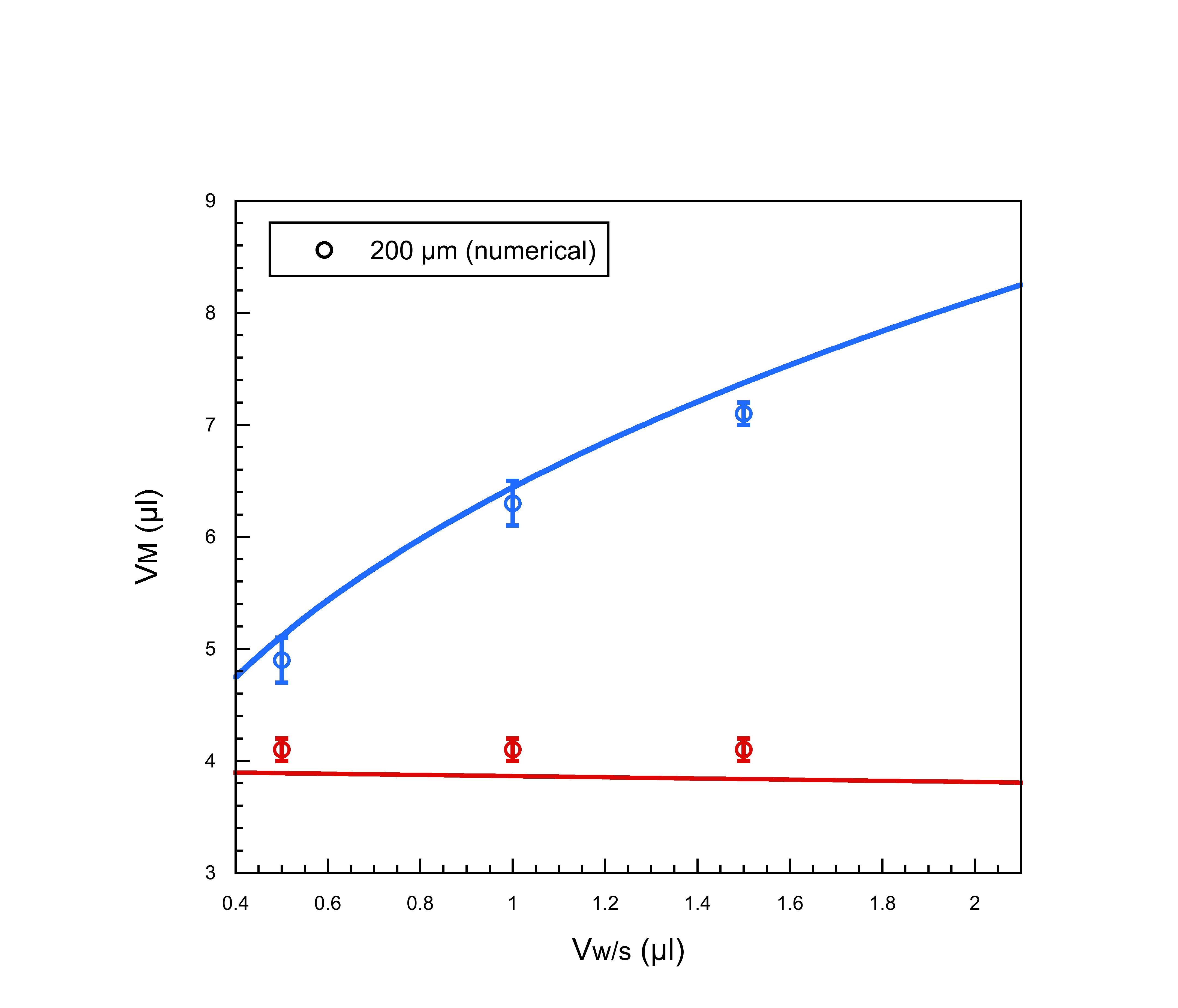}
\vskip -0.2 cm
\caption{The maximum volume determined by simulations as a function of the inner volume. Here, we considered $200 \ \rm{\mu m}$ fibers. We compare these numerical data with the model proposed for both pure and soapy water and represented by the blue and the red lines. It appears that the behavior of the system is well predicted by the simulations and that it is coherent with our models.}
\label{simu}
\end{center}
\end{figure}

These dissimilar behaviors are caused by the absence or the presence of a quadruple contact line. As pure water forms a core inside oil with two types of contact lines, the inner droplet plays the role of a portion of the fiber with a larger diameter. The water droplet acts like a defect of the fiber with its specific surface tension. The characteristic length is no longer the fiber diameter, as expected for a single droplet on a fiber \cite{lorenceau, lor}, but the inner droplet diameter which is much larger. So, larger volumes of oil can be hold. For soapy water, the inner drop is spread and therefore the characteristic length is still the fiber diameter. The whole system behaves as if it was made of only one component droplet.

For pure water, the oil droplet hangs around the water droplet like it would be hanging at the end of a capillary. As proposed in an earlier work \cite{quere}, it is possible to find the maximum volume for a droplet $V_M$ hanging from a capillary,
\begin{eqnarray} 
V_M = \frac{2 \pi c \gamma_{A/o} r_c}{\rho_o g}
\label{cap}
\end{eqnarray}
where $r_c$ is the radius of the capillary and $c$ a corrective factor which is about $0.7$. By using the radius of the water droplet as the radius of the capillary ($r_c = \sqrt[3]{(3/4\pi) V_w}$), Equation (\ref{cap}) can be rewritten as 
\begin{eqnarray} 
V_M = \zeta \frac{2 \pi c \gamma_{A/o} \sqrt[3]{(3/4\pi) V_w}}{\rho_o g}
\label{capbis}
\end{eqnarray}
and it is possible to fit the data with this new equation with a shape factor $\zeta$ which accounts for the fact that the oil droplet is hanging on the water droplet instead of a capillary. Therefore, the shape factor is not affected by the fiber diameter. A single shape factor is calculated and it is found to be equal to $0.81$. This fit is drawn in Figure \ref{deta} and is in a excellent agreement with our data. In this case, the fiber diameter does not influence the maximum volume. Indeed, the characteristic length is the inner droplet diameter instead of the fiber diameter. Therefore, as only the inner volume matters, the data are superimposed for the three different diameters. For small volumes of water, we noticed, both experimentally and numerically, that the maximal volume tends to the one found for soapy water. For these small volumes, the water droplet is so tiny that the oil droplet wraps not only the water droplet but also the fiber and therefore, the contact between the oil and the fiber can not be neglected anymore. The shape is, in that case, similar to the shape of the compound droplet with soapy water. It is therefore coherent that the pure water data converges towards the soapy water data. 

We performed simulations for compound droplets hanging on a fiber with a $200  \ \rm{\mu m}$ diameter. For each inner volume, the oil volume was progressively increased until the system detaches from the fiber. Figure \ref{simu} represents the numerical results. The maximal volume, represented by a circle, is therefore defined as the mean between the larger volume for which the droplet does not detach and the smaller volume for which the droplet detaches. The error bars join these two extreme volumes. One can notice that these data are in good agreement with the experimental results shown in Figure \ref{deta} as well as the analytical predictions given by Equation (\ref{capbis}). The simulation parameters are given in Table \ref{table:simulation_parameters}. For clarity, we specify the resulting contact angles instead of the set of used surface tensions. The densities for the different phases are not listed, since we used the same values as in the experiments (see Section II).
			\begin{table}[th!]
			\caption{Simulation parameters for the considered systems. The parameters $m$, $\gabd$ and $\epsilon$ are dimensionless.}
			% title of table
			\centering
			% used for centering table
			\begin{tabular}{p{3.5cm} c c c c c c c c}
			% centered columns (4 columns)
			\hline
			%inserts double horizontal lines
			System & $\alpha_w$ & $\alpha_s$ & $\beta$ & $\theta_w$ & $\theta_s$ & $m$ & $\gabd$ & $\epsilon$ \\ [0.5ex]
			% inserts table
			%heading
			\hline
			% inserts single horizontal line
			Pure water/oil & $60^\circ$ & $-$ & $25^\circ$& $62^\circ$ & $-$ & $2$ & $12$ & $4$\\
			% inserts body of the table
			Soapy water/oil & $-$ & $20^\circ$& $25^\circ$ & $-$ & $0^\circ$ & $0.5$ & $5$ & $4$\\
			\hline
			%inserts single line
			\end{tabular}
			\label{table:simulation_parameters}
			% is used to refer to this table in the text
			\end{table}

For soapy water, it seems that the maximal global volume is independent of the volume of the soapy water droplet. This means that only the sum of the water volume and oil volume matters. Thus, the system behaves like a droplet made of only one component due to the presence of the quadruple contact line. The detachment of  droplets from fibers has been studied \cite{mull, lorenceau, lor}. Lorenceau \textit{et al.} \cite{lorenceau, lor} developed a model for the detachment of a single droplet on a fiber and the maximal volume is given by
\begin{gather}
V_M = \frac{4 \pi  \gamma_{A/o} r_f }{\rho_o g}
\end{gather}
where $V_M$ is the maximum volume of the droplet and $r_f$ is the radius of the fiber. 
We can use this expression and replace $\rho_o$ with $(\rho_s V_s + \rho_o V_o)/(V_s+V_o)$ in order to take account of the difference in density of water and oil. Therefore, the maximum volume of a compound droplet containing soapy water is given by 
\begin{gather}
V_M = \xi \frac{4 \pi  \gamma_{A/o} r_f }{\rho_o g} - \frac{(\rho_s-\rho_o)}{\rho_o} V_{s}
\label{cap2}
\end{gather}
with a shape factor $\xi$ due to the fact that the droplet is made of two components. This expression is little dependent on the inner volume, $V_s$. Thus, the associate curve is a line with a slight slope as it is shown in Figure \ref{deta}. Moreover, the maximal volume depends on the fiber diameter, as it can be seen from both the data and the model. So, the larger the fiber, the larger the maximal volume. The shape factor has to be calculated for each fiber diameter. Indeed, as the maximal volume depends on $r_f$, the shape factor also depends on the fiber diameter. As a result, we find that $\xi = 0.77, 0.71$ and $0.68$ for $r_f = 200, 280$ and $350  \ \rm{\mu m}$, respectively. It proves that soapy-water oil droplets behave as simple systems thanks to the quadruple line. Once again, simulations have been performed and the maximal volumes obtained for soapy water are in good agreement the analytical predictions. Thus, simulations confirm not only the tendency observed experimentally, but also the model given by Equation (\ref{cap2}). Moreover, it can be seen, from the simulations, that the contact lines have merged in the vicinity of the fiber and that they move together along it (see supporting information).

%%%%%%%%%%%%%%%%%%%%%%%%%%%%%%%%%%%%%%%%%%% CONCLUSION
\vskip 0.4cm
\section{Conclusions}
In summary, we have shown that oil encapsulates water, also on fibers. Moreover, depending on the nature of the inner droplet, the system may act differently. For pure water, the triple lines remain separated. Whereas for soapy water, triple lines merge to form a quadruple line along the fiber. This phenomenon modifies the shape of the compound droplets. We have also shown that close to the detachment, the difference between pure and soapy water is enhanced. Both experimental and numerical results show that, in the case of pure water, the oil is hanging from the water droplet and the detachment only depends on the inner droplet diameter. Whereas, for soapy water, the oil moves with the soapy water along the fiber and therefore the detachment process is controlled by the fiber diameter.

Future works concern the dynamical properties of compound droplets on fibers. The fundamental question is how the droplet motion is affected by the nature and the size of the core droplet. Moreover, more complex systems, such as microemulsions, evaporation and condensation of droplets on fibers have to be investigated.

%%%%%%%%%%%%%%%%%%%%%%%%%%%%%%%%%%%%%%%%%%%
\vskip 0.4cm
{\noindent \bf \color{NavyBlue} Acknowledgments} F.Weyer is financially supported by an FNRS grant. This work is also supported by the FRFC 2.4504.12. Parts of the work from the Karlsruhe team have been achieved through funding by the German Research Foundation within the grant NE 822/20-1, which we gratefully acknowledge. 

\vskip 0.4cm
{\noindent \bf \color{NavyBlue} Supporting Information} Supporting Information Available: Two movies are presented and show the simulations of the detachment process for both pure and soapy water compound droplets . This material is available free of charge via the internet at http://pubs.acs.org.

%%%%%%%%%%%%%%%%%%%%%%%%%%%%%%%%%%%%%%%%%%%

\end{document}